\documentclass[aps,pra,twocolumn,nopacs,floatfix]{revtex4-2}
\usepackage[utf8]{inputenc}
\usepackage{textcomp}
\usepackage{epsfig}
\usepackage{graphicx} 
\usepackage{dcolumn}
\usepackage{amsthm,amsmath}
\usepackage{comment}
\usepackage[colorlinks=true, allcolors=blue]{hyperref}
\usepackage{xcolor}
\usepackage{colortbl}
\usepackage{multirow}
\usepackage{bigdelim}
\usepackage{colortbl}
\usepackage{mathtools}
\usepackage{orcidlink}
\usepackage{amsmath}
\usepackage[normalem]{ulem}
\usepackage{bm}
\usepackage[T1]{fontenc}
\usepackage{orcidlink}
\usepackage{graphicx}% Include figure files
\usepackage{dcolumn}% Align table columns on decimal point
\usepackage{bm}% bold math

\begin{document}

%\preprint{APS/123-QED}

\title{Investigating Role of Electron Correlation Effects via Triple Excitations for Precise Evaluation of Energies and Hyperfine Structure Constants in $^{23}$Na}

\author{$^{a,b}$Vaibhav Katyal\orcidlink{0000-0002-7717-8558}}
\email{vaibhavkatyal@prl.res.in}
\author{$^a$B. K. Sahoo\orcidlink{0000-0003-4397-7965}}
\email{bijaya@prl.res.in}

\affiliation{
$^a$Atomic, Molecular and Optical Physics Division, Physical Research Laboratory, Navrangpura, Ahmedabad 380009, India}  
\affiliation{
$^b$Indian Institute of Technology Gandhinagar, Palaj, Gandhinagar 382355, India
}

\begin{abstract}
Accurate determination of hyperfine structure constants in atomic systems provides important insight into the interplay of electron correlation and relativistic effects in the nuclear region. Although sodium (Na) is a relatively light atom, previous all-order relativistic many-body calculations of the magnetic dipole hyperfine constants for the low-lying states of $^{23}$Na show noticeable discrepancies with experiment. To address this, we calculate the ionization potentials and hyperfine structure constants of $^{23}$Na using relativistic coupled-cluster theory with explicit inclusion of triple excitations. We further incorporate corrections from the Breit interaction, quantum electrodynamics, and the Bohr–Weisskopf (BW) effect. Results from lower-order methods are also presented to assess the importance of different physical contributions across states. Our calculations demonstrate that contributions from the lower-order relativistic and BW effects play almost similar roles with the electron correlation effects, including triple excitations, and are essential for reconciling theoretical predictions with experimental observations. This study can also serve as a useful guide for understanding the role of triples in heavier alkali systems.
\end{abstract}

\maketitle

\section{Introduction}

Understanding the role of electron correlation effects in many-body systems is of fundamental importance in quantum chemistry and atomic physics. Among various multi-electron systems, atomic systems provide a relatively simpler platform for investigating electron correlation phenomena because of their well-defined electronic structures and the availability of highly accurate theoretical methods \cite{shavitt2009many,grant2007relativistic,lindgren2012atomic}. Consequently, atomic properties can often be calculated with significantly higher precision than those of more complex molecular or condensed-matter systems. Nevertheless, despite significant progress in theory and experiment, discrepancies between calculated and measured atomic properties still persist in some cases, often due to subtle many-body interactions and inadequate treatments of electron correlation. Importantly, the nature and magnitude of correlation effects depend not only on the size of the system but also on the electronic configuration of the state, which determines the relative contributions of core, core-valence, and valence correlations. Therefore, detailed studies of correlation effects in carefully chosen atomic systems are essential for developing reliable many-body theories.

Among all elements in the periodic table, 
alkali atoms are of particular interest because of their simple electronic structure and broad applications in precision spectroscopy \cite{hansch2006nobel}, atomic clocks \cite{oac}, quantum information \cite{saffman2010quantum}, and tests of fundamental physics \cite{ginges2004violations}. Furthermore, their low-lying states can be described as a single valence electron outside a closed-shell core, making them excellent testbeds for investigating many-body effects and benchmarking sophisticated many-body approaches. Since the dominant correlation contributions arise from valence electrons while core correlations remain relatively small, alkali atoms offer a unique opportunity to isolate and quantify different classes of electron correlation effects. Consequently, studies of alkali atoms provide valuable insights into many-body interactions and help improve high-precision theoretical approaches applicable across a wide range of quantum systems.

Sodium ($^{23}$Na) has long served as an important benchmark system for studies of hyperfine interactions because of its simple single-valence-electron structure and the availability of highly precise spectroscopic measurements for several low-lying states \cite{beckmann1974precision, volz1996precision, van1994measurement, happer1974atomic, burghardt19883}. An early calculation of the magnetic-dipole hyperfine-structure constant ($A_{hf}$) in $^{23}$Na, based on non-relativistic coupled-cluster theory with single and double excitations (CCSD) supplemented by relativistic corrections estimated at the Dirac–Hartree–Fock (DHF) level \cite{salomonson1991coupled}, showed good agreement with experiment for the ground state. That study explicitly demonstrated the importance of electron-correlation effects, particularly core polarization (CP) and pair-correlation (PC) contributions, in determining $A_{hf}$ values. Subsequently, calculations employing large basis sets within both the non-relativistic multi-configuration Hartree–Fock (MCHF) and relativistic multi-configuration Dirac–Hartree–Fock (MCDHF) frameworks \cite{jonsson1996large} confirmed the significant role of relativistic effects. Later, a relativistic coupled-cluster (RCC) calculation retaining only the linearized single and double excitation terms (SD method) yielded a ground-state $A_{hf}$ value slightly larger than the experimental result \cite{safronova1998relativistic}. However, when nonlinear terms were included through the full RCC singles-and-doubles (RCCSD) method, the predicted value became substantially lower than experiment \cite{pal2007relativistic}. Interestingly, the experimental value lies between the SD and RCCSD predictions. A similar trend was observed in $^{7}$Li, where the discrepancy between theory and experiment was largely resolved by incorporating triple excitations \cite{derevianko2008convergence}. However, in $^{7}$Li, triple excitations arise only through valence correlations because its closed core contains only two electrons. For $^{23}$Na perturbative inclusion of triple excitations was found to nearly cancel the nonlinear RCCSD contributions in the calculated $A_{hf}$ alues \cite{porsev2006triple}.

Despite these extensive investigations, the individual roles of higher-order electron-correlation effects in determining the hyperfine structure of $^{23}$Na  remain incompletely understood. Although Na is a relatively light atomic system, for which higher-order correlation effects are generally expected to be small, a quantitative understanding of these contributions is essential for establishing the accuracy and reliability of many-body methods. Such an understanding is important not only for achieving precise hyperfine-structure calculations in Na but also for developing robust theoretical approaches that can be confidently extended to heavier alkali-metal atoms, where electron-correlation effects are substantially more pronounced.

Recently, renewed measurements of the $A_{hf}$ of the $3s^2S_{1/2}$ and $3p^2P_{1/2}$ states in both the $^{21}$Na and $^{23}$Na isotopes have been reported \cite{won2026experimental}. To explain results from these high-precision measurements, relativistic coupled-cluster theory with single, double, and triple excitations (RCCSDT method) was employed. This combined theoretical and experimental study revealed several interesting features. In particular, the RCCSD values of $A_{hf}$ for these states were found to be in close agreement with the earlier results reported in Ref.~\cite{porsev2006triple}. However, inclusion of the valence triple excitations within the RCCSD framework (RCCSDTv) increased the calculated $A_{hf}$ values beyond the experimental results. The subsequent inclusion of full triple excitations through the RCCSDT method though shifted the ground-state value little closer to the measurement but there was still significant gap between them. Interestingly, when additional relativistic and nuclear corrections arising from the Breit interaction, vacuum-polarization (VP), self-energy (SE), and Bohr--Weisskopf (BW) effects were taken into account, the theoretical results came into close agreement with the measured values. These findings demonstrated that, although the individual Breit, quantum electrodynamic (QED), and BW contributions are relatively small, their combined effect is comparable to that of the higher-order electron-correlation corrections and is essential for achieving theoretical accuracy.

The interplay between electron correlation effects at different level of approximation in the many-body method and relativistic effects for accurate evaluation of the ground state $A_{hf}$ value motivated us to investigate whether similar trends persist in other excited states of $^{23}$Na. Further, we would like to examine the roles of triple excitations and relativistic corrections in the determination of ionization potentials (IPs) and electric-quadrupole ($B_{hf}$) hyperfine-structure constants of $^{23}$Na. To gain deeper insight into the propagation of electron-correlation effects in both $A_{hf}$ and $B_{hf}$ of $^{23}$Na across different levels of many-body approximations, we also evaluate these quantities using several lower-order approaches, including second-order relativistic many-body perturbation theory [RMBPT(2) method], third-order relativistic many-body perturbation theory [RMBPT(3) method], the random-phase approximation (RPA), the Br\"uckner-orbital (BO) approach, the structural-radiation (SR) correction approach, and the normalization (Nm) correction approach. This systematic comparison enables us to assess the relative importance of various correlation mechanisms and higher-order effects in hyperfine-structure calculations of $^{23}$Na. To provide a comprehensive analysis, we have undertaken as many as 11 low-lying states of Na in the present work. 

\section{Theory of hyperfine interaction}

The noncentral Hamiltonian describing hyperfine interactions between electrons and atomic nucleus can be expressed as \cite{schwartz1955hfs}
\begin{equation}\label{HFS_eq}
H_{\text{hfs}} = \sum_{k=1} M_n^{(k)}(I) \cdot T_e^{(k)}(J) , 
\end{equation}
where $M_n^{(k)}(I)$ and $T_e^{(k)}(J)$ are spherical tensor operators of rank $k$ in the nuclear and electronic coordinate spaces, respectively. Here $I$ and $J$ are the net nuclear and electron angular momenta, respectively, which produces hyperfine angular momentum $F = J \oplus I$ through interaction. In the above expression, $k=1$ term corresponds to magnetic dipole (M1) interaction, $k=2$ term stands for the electric quadrupole (E2) interaction and so on. 

Since hyperfine interactions are relatively weak compared to other electromagnetic interactions in an atomic system, their contributions are typically treated within perturbation theory. 
Assuming the unperturbed wave function
\begin{equation}
\left|(I J) F M_F\right\rangle \equiv \left|I M_I; J M_J\right\rangle ,
\end{equation}
where $M$s are the respective azimuthal quantum numbers, the first-order correction to the atomic energy levels due to the hyperfine interactions can be expressed as 
\begin{eqnarray}
   W_{F, J}^{(1)} &=& \langle (I J) F M_F|H_{hfs}|(I J) F M_F\rangle \nonumber \\ 
    &=& (-1)^{I+J+F} \sum_{k} \left\{ \begin{matrix}
J & I & F \\
I & J & k \\
\end{matrix} \right\} \nonumber \\ && \times \left \langle I \left \| M_n^{( k)} \right\| I \right \rangle \left \langle J \left \| T_e^{(k)} \right \| J \right \rangle .
\end{eqnarray}
It is obvious from the above expression that $ W_{F, J}^{(1)}$ is nonzero only when the reduced matrix elements $\left \langle I \left \| M_n^{( k)} \right\| I \right \rangle$, $ \left \langle J \left \| T_e^{(k)} \right \| J \right \rangle$ and $F$ are nonzero. As mentioned before, we only retain the $k=1$ and $k=2$ terms for our investigation due to their dominance contributions. The nuclear operators are defined as 
\begin{eqnarray}
 \left \langle I I | M_n^{( 1)} | I I \right \rangle = \left ( \begin{matrix}
I & 1 & I \\
-I & 0 & I \\
\end{matrix} \right )  \left \langle I \left \| M_n^{( 1)} \right\| I \right \rangle = \mu_I
\end{eqnarray}
and 
\begin{eqnarray}
 \left \langle I I | M_n^{( 2)} | I I \right \rangle = \left ( \begin{matrix}
I & 2 & I \\
-I & 0 & I \\
\end{matrix} \right )  \left \langle I \left \| M_n^{( 2)} \right\| I \right \rangle = \frac{1}{2} Q_I ,
\end{eqnarray}
where $\mu_I$ and $Q_I$ are the nuclear M1 and E2 hyperfine structure constants, respectively. The electronic component operators are given by 
 \cite{schwartz1955hfs}
\begin{eqnarray}
 T_e^{(1)} &=& -ie \sum_{q=-1}^1 \sum_j \sqrt{8 \pi / 3} r_j^{-2} \alpha_{\mathbf{j}} \cdot \mathbf{Y}_{\mathbf{1 q}}^{(\mathbf{( 1 )}}\left(\mathbf{r}_{\mathbf{j}}\right)
 \end{eqnarray}
 and 
 \begin{eqnarray}
 T_e^{(2)} &=& - ie \sum_{q=-2}^2 \sum_j r_j^{-3} C_q^{(2)}\left(r_j\right),
\end{eqnarray}
where $C_q^{(k)}$ is the Racah operator of rank $k$ and $\alpha$ is the Dirac operator. 

Therefore, the net first-order hyperfine interaction energy correction is given by 
\begin{equation}
W_{F,J}^{(1)} \simeq W_{F,J}^{M1}+W_{F,J}^{E2},
\end{equation}
where $W_{F,J}^{M1}$ and $W_{F,J}^{E2}$ denote the contributions arising from the $M1$ and $E2$ interactions, respectively. Conventionally, these contributions are mathematically expressed as
\begin{eqnarray}
W_{F,J}^{M1} &=& A_{hf} K 
\end{eqnarray}
and
\begin{eqnarray}
W_{F,J}^{E2} &= B_{hf} \, \frac{ 3( K^2  +\frac{3}{2} K -I(I+1)J(J+1)}{2I(2I-1)J(2J-1)}.
\end{eqnarray}
where $K=\frac{1}{2} \left [ F(F+1)-J(J+1) - I(I+1)\right ]$.

It follows the definitions of the M1 and E2 hyperfine structure constants as
\begin{eqnarray}
A_{hf} &= \frac{\mu_Ng_I}{\sqrt{J(J+1)(2J+1)}}\, \left\langle J  ||T_e^{(1)} || J\right\rangle
\end{eqnarray}
and
\begin{eqnarray}
B_{hf} &= 2eQ_I\left\{\frac{8J(2J-1)}{(2J+1)(2J+1)(2J+3)}\right\}  \left\langle J  ||T_e^{(1)} || J\right\rangle,
\end{eqnarray}
where $g_I=\mu_I/I$ is the nuclear gyromagnetic ratio, and $\left\langle J  ||T_e^{(1)} || J\right\rangle$ and $\left\langle J  ||T_e^{(1)} || J\right\rangle$ are the M1 and E2 electronic reduced matrix elements, respectively.

\begin{table*}[t] 
\centering
\caption{Calculated ionization potentials (IPs), in cm$^{-1}$, at different levels of approximations in the methods. The final recommended values by combining RCCSDT results and other relativistic corrections from the present calculation, along with their uncertainties, are also given and compared with the available NIST data \cite{kramida2018nist}.}
\begin{tabular}{l rrrr r|rrr| rr}
\hline \hline
State  & \multicolumn{1}{c}{DHF}  & \multicolumn{1}{c}{RMBPT(2)} &  \multicolumn{1}{c}{RCCSD}  &  \multicolumn{1}{c}{RCCSDTv} & RCCSDT  & +Breit & +VP & +SE& \multicolumn{1}{c}{Final}  & \multicolumn{1}{c}{NIST}\\
\hline \\
$3S_{1/2}$ & 39951.56  & 41220.22 & 41358.62 & 41456.37 & 41437.63 &  $-1.86$ & 0.15 & $-3.72$ & 41432(20)  & 41449.45   \\
$3P_{1/2}$ & 24030.34 & 24413.75 & 24464.95  & 24496.07 & 24492.36 & $-1.43$ & $-0.01$ & $0.09$ &  24491(5) & 24493.28    \\
$3P_{3/2}$ & 24014.12 & 24395.62 & 24446.56 & 24477.55 & 24473.49  & $-0.32$ & $-0.01$ & $0.01$ & 24473(4) & 24476.08  \\
$4S_{1/2}$ & 15398.78  & 15668.93 & 15691.34 & 15710.80 & 15706.98 & $-0.46$ & 0.04 & $-0.83$ & 15706(10)  & 15709.26 \\
$3D_{3/2}$ & 12217.34  & 12266.70  & 12273.62 & 12276.74 & 12276.27  & 0.06 & $\sim 0$ & $\sim 0$ & 12276(1) & 12276.56 \\
$3D_{5/2}$ & 12217.38  & 12266.74  & 12273.66 & 12276.78 & 12276.32 & 0.04 & $\sim 0$ & $\sim 0$ & 12276(1) & 12276.61\\
$4P_{1/2}$ & 11047.79 & 11160.55 & 11173.94 & 11183.17 & 11182.04 & $-0.49$ & $\sim 0$ & 0.06 & 11181(2) & 11182.26    \\
$4P_{3/2}$ & 11042.38 & 11154.61 & 11167.95 & 11177.14 & 11175.90 & $-0.11$ & $\sim 0$ & 0.04 & 11175(2) & 11176.67  \\
$5S_{1/2}$ & 8132.62  & 8233.69 & 8241.29 & 8248.41 & 8247.00 & $-0.18$ & 0.01 & $-0.32$ & 8246(3)  & 8248.58 \\
$4D_{3/2}$ & 6872.50  & 6895.74 & 6898.93 & 6900.43 & 6900.19 & 0.03 & $\sim 0$ & $\sim 0$  & 6900(1)  & 6900.49\\
$4D_{5/2}$ & 6872.55 & 6895.77  & 6898.96 & 6900.47 & 6900.22 & 0.02 & $\sim 0$ & $\sim 0$ & 6900(1)  & 6900.52 \\
\hline \hline
\end{tabular}
\label{tab2}
\end{table*}

\section{Computational details}

In this section, we discuss different approximations made both in the atomic Hamiltonian and many-body methods to carry out calculations of wave functions, energies and hyperfine structure constants. We also discuss briefly about the basis functions used.

\subsection{Approximations in the Hamiltonian}

The Dirac-Coulomb (DC) atomic Hamiltonian employed in the present case is given in atomic units (a.u.) by
\begin{eqnarray}
H_{DC} &=& \sum_i \Lambda_i^+ \left[c\boldsymbol{\alpha}\cdot \mathbf{p}_i + (\beta-1)c^2 + V_N(r_i)\right] \Lambda_i^+ \nonumber \\ && + \sum_{i<j} \Lambda_i^+ \Lambda_j^+ \frac{1}{r_{ij}} \Lambda_j^+ \Lambda_i^+,
\end{eqnarray}
where $\Lambda^+$ operators denote projection on the electronic component, $\boldsymbol{\alpha}$ and $\beta$ are the usual Dirac matrices, $V_N(r)$ is the nuclear potential generated from a finite charge distribution, and the last term represents the electron-electron interaction due to exchange of longitudinal photon. Contributions from the transverse photon exchange is estimated by considering the frequency-independent Breit interaction potential, given in a.u. by 
\begin{equation}
V_B(r_{ij})=-\Lambda_i^+ \Lambda_j^+ \left[ \frac{\boldsymbol{\alpha}_i\cdot \boldsymbol{\alpha}_j} {2r_{ij}} + \frac{(\boldsymbol{\alpha}_i\cdot \mathbf{r}_{ij})
(\boldsymbol{\alpha}_j\cdot \mathbf{r}_{ij})} {2r_{ij}^3}\right]\Lambda_j^+ \Lambda_i^+  . \nonumber 
\end{equation}

For precise results, we further evaluate QED corrections to the hyperfine constants. The VP contributions are incorporated through the local Uehling and Wichmann-Kroll (WK) potentials. The Uehling potential gives dominant vacuum-polarization contribution, which for a finite nucleus is given by
\begin{eqnarray}
V_U(r) &=& -\frac{2 }{3 } \frac{\alpha_e^2}{r} \int_0^{\infty} dx \, x \, \rho_N(x) \, \int_1^\infty dt\, \sqrt{t^2-1} \nonumber \\
 && \times \left(\frac{1}{t^3}+\frac{1}{2t^5}\right) \left [ e^{-2mct|r-x|} - e^{-2mct|r+x|}  \right ], \ \ \ \ \
\end{eqnarray}
where $\alpha_e$ is the fine-structure constant, $m$ is the electron mass and $\rho_N(r)$ is the nuclear density normalized to atomic number $Z$. We have assumed Fermi-charge distribution within the atomic nucleus to define $V_N(r)$ and $\rho_N(r)$. The WK potential for a finite nucleus is defined as 
\begin{equation}
V_{WK}(r)= \frac{0.368Z^2}{9 \pi c^3 [1+(1.62cr)^4] } \, \rho_N(r) .
\end{equation}

While defining the VP potentials are relatively simpler, treatment of SE contributions are considerably more challenging owing to its intrinsically nonlocal and energy-dependent nature. The dominant SE contribution is evaluated using the Flambaum-Ginges radiative potential approach (FGRP) \cite{flambaum2005radiative, sahoo2021benchmarking}, in which a local radiative potential is added to the atomic Hamiltonian to simulate one-loop bound-state QED corrections.

\subsection{Many-body methods}

We intend to calculate properties of the $3S$, $3P_{1/2;3/2}$, $3D_{3/2;5/2}$ $4S$, $5S$, $4P_{1/2;3/2}$, $4D_{3/2;5/2}$ states of Na in the present study. These states share a common closed-core configuration, $[2p^6]$, and differ only in the valence orbital $v$. In order to produce all the states with a common reference state, we obtain first the mean-field reference state for the $[2p^6]$ configuration, $|\Phi_0 \rangle$, using the $H_{DHF}$ Hamiltonian in the DHF method. Subsequently, the respective valence orbital of an state was appended to obtain the desired electronic configuration of the state. The residual interactions ($V_{res}$) neglected at the DHF method were accounted in two steps. First, the correlations among the electrons within the closed-core were accounted and then, the interactions involving the valence electron were included.

To describe the above procedure theoretically, the atomic Hamiltonian in general is defined as $H_{at}=H_{DHF} + V_{res}$ and the exact wave function of the closed-core, $|\Psi_0 \rangle$, is obtained by employing a wave operator $\Omega_0$ on $|\Phi_0 \rangle$; i.e. $|\Psi_0 \rangle = \Omega_0 |\Phi_0 \rangle$. Then, the exact wave function of a state with the valence orbital $v$ is produced by defining another wave operator $\Omega_v$ such that $|\Psi_v \rangle = \left ( \Omega_0 + \Omega_v \right ) |\Phi_v \rangle $, where $|\Phi_v \rangle = a_v^{\dagger} |\Phi_0 \rangle$. The amplitudes of the wave operators due to $V_{res}$ can be obtained by solving the following Bloch equations \cite{katyal2025testing}
\begin{eqnarray}
 \langle \Phi_0^* | [\Omega_0, H_{DHF} ] |\Phi_0 \rangle &=& \langle \Phi_0^* | V_{res} \Omega_0 |\Phi_0 \rangle
 \label{bl00}
\end{eqnarray}
and
\begin{eqnarray}
 \langle \Phi_v^* | [\Omega_v, H_{DHF} ] |\Phi_v \rangle &=& \langle \Phi_v^* | V_{res} (\Omega_0 + \Omega_v ) |\Phi_v \rangle  \nonumber \\ && -
 \langle \Phi_v^* | \Omega_v |\Phi_v \rangle E_v  ,
 \label{bl01}
\end{eqnarray}
where $E_v$ is the energy corresponding to the state $|\Psi_v \rangle$ and is evaluated by $E_v = \langle \Phi_v | H_{at} \left ( \Omega_0 + \Omega_v \right )  | \Phi_v \rangle$. The above equations can be solved order-by-order sequentially in the RMBPT method by expressing
\begin{eqnarray}
\Omega_0 &=& \sum_k \Omega_0^{(k)}, \ \ 
\Omega_v = \sum_k \Omega_v^{(k)} \ \ \text{and} \ \ 
E_v = \sum_k E_v^{(k)} .  \ \ \ \
\end{eqnarray}
In an all-order perturbative method, the amplitudes of the wave operators and energies are obtained by solving them simultaneously in the iterative procedure. In the all-order RPA method, the wave operators are defined as
\begin{eqnarray}
\Omega_{0} =  \sum_{a,p} \Omega_a^p \ \ \ \text{and} \ \ \  \Omega_{v} = \sum_{p} \Omega_v^p ,
\end{eqnarray}
where 
\begin{eqnarray}
 \Omega_a^p &=& \sum_k \sum_{b,q} \left [ \frac{\langle pb|V_{res}| aq \rangle - \langle pb|V_{res}| qa \rangle}{\epsilon_a - \epsilon_p } \Omega_b^{q,(k-1)} \right. \nonumber \\
 && \left. + \Omega_b^{q,(k-1)\dagger} \frac{\langle pq|V_{res}| ab \rangle - \langle pq|V_{res}| ba \rangle}{\epsilon_a - \epsilon_p} \right ] .
\end{eqnarray}
In the above expressions, $a,b,\cdots$ and $p,q,\cdots$ denote for core and virtual DHF orbitals with their energies denoted by $\epsilon$. By construction, RPA wave operators account CP effects to all-orders through single excitations. However, it does not include the PC and SR corrections. To show their importance, we add PCs through BO and SR corrections in the RMBPT method at the RMBPT(3) approximation. Adding these corrections result in wave functions are not normalized, which are then accounted for by the normalization factors. Detailed discussions on these methods can be found in Ref. \cite{arup2024}.

\begin{table*}[t]
\caption{Calculated $A_{hf}$ constants of $ ^{23}$Na (in MHz), obtained using $g_I=\mu_I/I=1.4784371(6)$ \cite{stone2005table} at various levels of many-body approximations. Here, BO denotes for the Br\"uckner orbital contribution, SR represents for the structural radiation correction, Nm corresponds to the normalization correction in the RMBPT method, and $+\mathrm{BW}$ indicates the Bohr--Weisskopf correction. The final values are given after adding all corrections with the RCCSDT values. The final values with quoted uncertainties are compared with the values obtained using other theories and available precise measurements}
\resizebox{\textwidth}{!}{\begin{tabular}{lccccccccccc}
\hline \hline \\
Method & $3 S_{1 / 2}$ & $3 P_{1 / 2}$ & $3 P_{3 / 2}$ & $4 S_{1 / 2}$ & $3 D_{3 / 2}$ &  $3 D_{5 / 2}$ & $4 P_{1 / 2}$ & $4 P_{3 / 2}$ & $5 S_{1 / 2}$ &  $4 D_{3 / 2}$ & $4 D_{5 / 2}$ \\
\hline  & & & & & & & & & & \\
DHF & 623.97 & 63.43 & 12.60 & 150.56 & 0.59 & 0.25 & 20.98 & 4.17 & 58.16 &   0.25 & 0.11 \\
RMBPT(2) & 739.39 & 77.10 & 15.31 & 178.20 & 0.52 & 0.17 & 25.51 & 5.04 &  68.73 &  0.21 & 0.06 \\
RMBPT(3) & 827.17 & 85.03 & 16.96 & 194.01 & 0.54 & 0.18 & 27.81 & 5.53 & 74.21 &    0.22 & 0.07 \\
RPA & 767.64 & 82.33 & 18.01 & 184.79 & 0.50 & 0.10 & 27.22 & 5.92 &  71.26 &    0.20 & 0.02 \\
RPA+BO & 861.81 & 91.09 & 19.75 & 202.04 & 0.52 & 0.11 & 29.78 & 6.43 & 77.28 &    0.21 & 0.02\\
RPA+BO+SR & 853.90 & 89.90 & 18.80 & 200.16 & 0.54 & 0.12 & 29.38 & 6.12 & 76.56 &  0.23 & 0.03\\
RPA+BO+SR+Nm & 852.38 & 89.85 & 18.79 & 200.09 & 0.54 & 0.12 &  29.38 & 6.12 & 76.55 &   0.23 & 0.03\\
RCCSD & 881.85 & 93.27 & 18.47 & 203.67 & 0.54 & 0.12 & 30.19 & 5.94 & 77.71 &    0.22 & 0.03\\
RCCSDTv & 898.95 & 96.36 & 18.85 & 206.73 & 0.53 & 0.11 &31.12 & 6.05 &   78.78 &   0.22 & 0.02\\
RCCSDT  & 890.82 & 94.63 & 18.49 & 205.36 & 0.53 & 0.11 & 30.61 & 5.94 &  78.26 &   0.22 & 0.03 \\
\hline \\
$+$Breit & $-0.55$ & $-0.08$ & $-0.02$ & $-0.14$ & $\sim 0$ & $\sim 0$ & $-0.03$ & $-0.01$ &  $-0.05$ &  $\sim 0$ & $\sim 0$\\
$+$VP  & $-0.78$ & $-0.12$ & $-0.03$ & $-0.19$ & $\sim 0$ & $\sim 0$ &  $-0.04$ & $-0.01$ &  $-0.07$ &  $\sim 0$ & $\sim 0$  \\
$+$SE  & $-3.98$ & 0.01 & $\sim 0$ &  $-0.91$ & $\sim 0$ & $\sim 0$ & 0.02 & 0.01 & $-0.27$ &   $\sim 0$ & $\sim 0$ \\
$+$BW  & $-0.31$ & $\sim 0$ & $\sim 0$ & $-0.07$ & $\sim 0$ & $\sim 0$ & $\sim 0$ & $\sim 0$ & $-0.03$ &    $\sim 0$ & $\sim 0$ \\
\hline \\
Final & 885.2(2.0) & 94.44(1.0) & 18.44(50)  & 204.05(2.0) & 0.53(5) & 0.11(1) & 30.56(50) & 5.93(10) & 77.84(1.0) &  0.22(2) & 0.03(1)\\
Other theories & 882.2 \cite{jonsson1996large} & & & & & & & & &\\
 &  883.8 \cite{salomonson1991coupled} &  & & 202.2 & & & & & &   \\
Experiments & 885.8 \cite{beckmann1974precision} &  94.42 \cite{carlsson1992multi} & 18.79 \cite{volz1996precision} & 202(3)\cite{gupta1973hyperfine} & 0.53 \cite{burghardt19883} & 0.11 \cite{burghardt19883}  & &  & 78(5) \cite{eliav1994ionization} &\\
& 884.9(16) \cite{won2026experimental} & 94.44(13) \cite{van1994measurement} & 18.53(15)\cite{yei1993delayed} &  &   & & & 6.01(3) \cite{happer1974atomic}   & & \\
&  & 95.9(10) \cite{won2026experimental} \\
\hline
\hline
\end{tabular}}
\label{Ahf}
\end{table*}

It is obvious from the above discussions that approximated RMBPT method cannot include higher-order correlation effects as it is limited by the finite order of perturbation theory. Although the all-order RPA sums the CP contributions to infinite order, it neglects several other important correlation effects, mainly PC, SR, and their correlations. While some of these missing contributions can be incorporated separately through perturbative corrections, such an approach fails to account for the interplay among the different classes of many-body effects. In contrast, the RCC method accounts for RPA and all aforementioned non-RPA effects simultaneously to all-orders. In the RCC approach, we express
\begin{eqnarray}
\Omega_0 = e^T \ \ \ \text{and} \ \ \ \Omega_v=e^T S_v .
\end{eqnarray}
It follows the energy expression 
\begin{equation}
E_v = \langle \Phi_v | [H_{at} e^T (1+S_v)]_{conn} |\Phi_v\rangle,
\end{equation}
where subscript $conn$ means only the connected terms are included and by defining normal ordering for $H_{at}$ with respect to $|\Phi_0 \rangle$, we calculate $E_v$ as ionization potential of the valence electron than total energy of the state. In terms of level of excitations, the RCC operators are expressed as
\begin{eqnarray}
T &=& T_1 + T_2 + T_3 + \cdots \nonumber \\
  &=& \sum_{ap} t_a^p a_p^{\dagger} a_a + \frac{1}{2} \frac{1}{2!} \sum_{ab,pq} t_{ab}^{pq} a_p^{\dagger} a_q^{\dagger} a_b a_a \nonumber \\
  && + \frac{1}{4} \frac{1}{3!} \sum_{abd,pqr} t_{abd}^{pqr} a_p^{\dagger} a_q^{\dagger}a_r^{\dagger}  a_d a_b a_a  + \cdots
\end{eqnarray}
and
\begin{eqnarray}
S_v &=& S_{1v} + S_{2v} + S_{3v} + \cdots \nonumber \\
  &=& \sum_{p} s_v^p a_p^{\dagger} a_v + \frac{1}{2!} \sum_{b,pq} s_{b}^{pq} a_p^{\dagger} a_q^{\dagger} a_b a_v \nonumber \\
  && + \frac{1}{2} \frac{1}{3!} \sum_{bd,pqr} s_{bd}^{pqr} a_p^{\dagger} a_q^{\dagger}a_r^{\dagger}  a_d a_b a_v  + \cdots ,
\end{eqnarray}
where subscripts 1, 2, 3, etc stand for the singly, doubly and triply excitation configurations, respectively, and $t$ and $s$ denote amplitudes of the $T$ and $S_v$ wave operators, respectively. We have approximated the RCC theory to the RCCSD method by defining 
\begin{eqnarray}\label{rccsd}
T = T_1 + T_2 \ \ \ \text{and} \ \ \ S_v=S_{1v} + S_{2v} ,
\end{eqnarray}
to the RCCSDTv method by defining  
\begin{eqnarray}\label{rccsdtv}
T = T_1 + T_2 \ \ \ \text{and} \ \ \ S_v=S_{1v} + S_{2v} + S_{3v} 
\end{eqnarray}
and to the RCCSDT method by defining
\begin{eqnarray}\label{rccsdt}
T = T_1 + T_2 + T_3 \ \ \ \text{and} \ \ \ S_v=S_{1v} + S_{2v} + S_{3v} 
\end{eqnarray}
to analyze importance of the core and valence triple excitation configurations explicitly in our calculations.

After solving amplitudes of the RCC wave operators, we evaluate expectation value of an operator $O$ as
\begin{equation}\label{prop_exp}
\langle O \rangle = \frac{\langle\Phi_v|(1+S_v^\dagger)e^{T^\dagger} O e^T(1+S_v)|\Phi_v\rangle}{\langle\Phi_v|(1+S_v^\dagger)e^{T^\dagger}e^T(1+S_v)|\Phi_v\rangle}.
\end{equation}
It can be noticed from the above expression that normalization factor (Nm) of the wave function has been included. As demonstrated in the earlier works \cite{sahoo2007enhanced, sur2005comparative, sahoo2004ab}, contributions from $OS_{1v}+c.c.$, with $c.c.$ denoting complex conjugate, give rise to PC effects to all-orders and contributions from $OS_{2v}+c.c.$ correspond to CP effects to all-orders, while the remaining terms correspond to SR and correlations among PC, CP and SR effects. 

\begin{figure}[t]
\centering
\includegraphics[width=8cm, height=6cm]{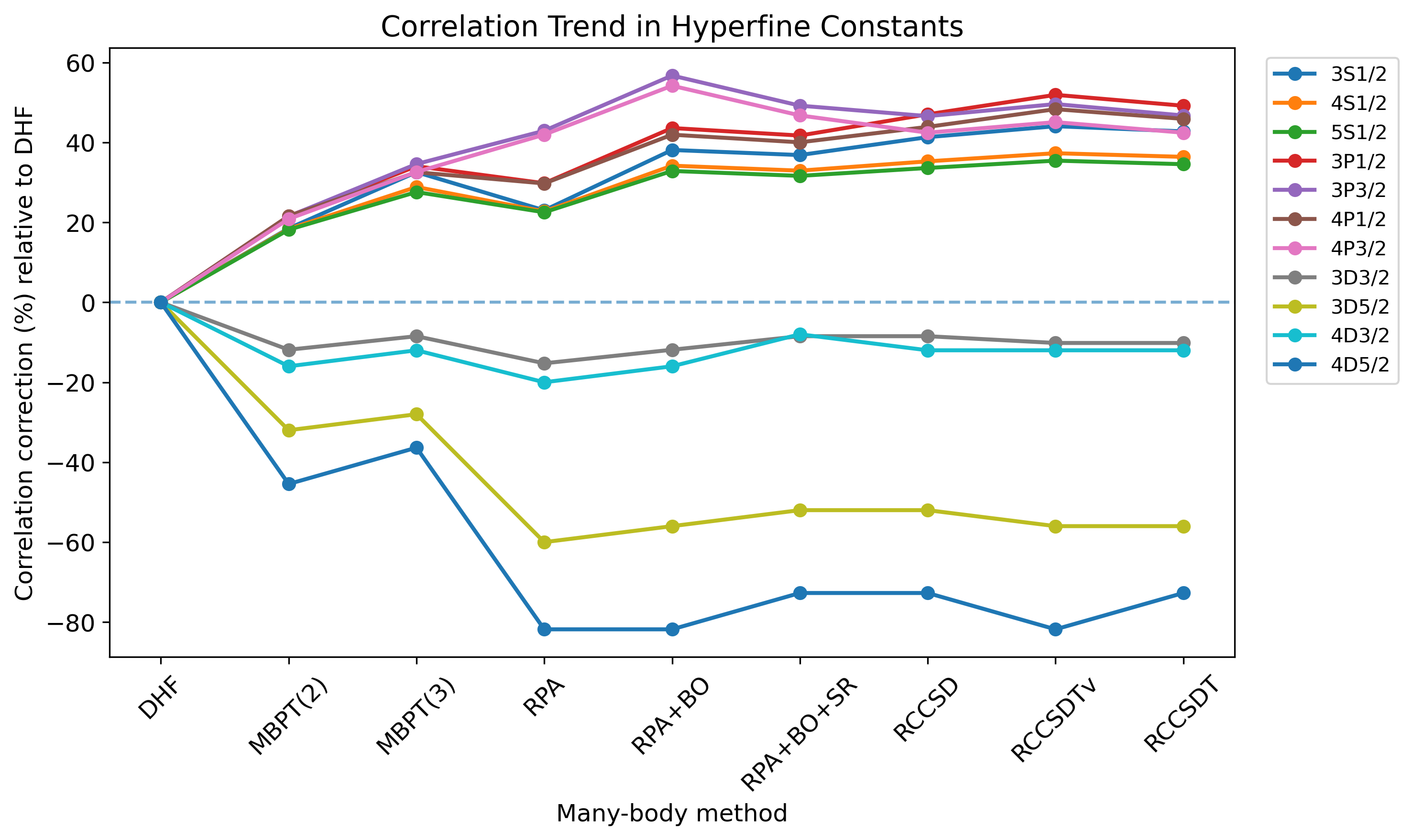}
\caption{Correlation trends from lower- to higher-order many-bdy methods in the calculated $A_{hf}$ values (in MHz) of the considered states in $^{23}$Na. The plotted quantities represent the percentage corrections with respect to the DHF values as progressively through the higher-order many-body methods.}
\label{corr_trend}
\end{figure}

\subsection{Basis functions}

Beside employing a suitable many-body method, the choice of appropriate basis set is equally important for obtaining accurate wave functions and, consequently, reliable atomic properties. Among the commonly used basis functions, GTOs are known to describe mean-field single particle wave functions more accurately within atomic nucleus. Using these basis functions, the radial functions of both the large and small components are generated according to a geometric progression, given by
\begin{eqnarray}
 f(r) = \sum_k C_k N_k e^{-\alpha_0 \beta^{k-1} r^2} , 
\end{eqnarray}
where $C_k$ is the linear coefficient, $N_k$ is the normalization factor, and $\alpha_0$ and $\beta$ are two arbitrary parameters. The $\alpha_0$ and $\beta$ are considered suitably in the even-tempered prescription so that a balanced and systematic description of both the near-nuclear and asymptotic regions of the electronic wave functions are taken care. We also impose kinetic balance condition between the large and small components of the DHF orbitals. We have chosen up to 40 GTOs for orbitals belonging to each $l$ with maximum value $l=6$ (i.e. up to $i$- symmetry), which is a considerably large space to consider in our RCC methods to account for electron correlation effects for Na. We also analyze contributions from high-lying orbitals through the RMBPT(2) method and accounted them in the uncertainty estimations.

To account for magnetization distribution within the atomic nucleus to estimate the BW effect in the Fermi-charge model, we define a function \cite{sahoo2025precise}. 
\begin{align*}
F_{mg}(r)= & \frac{1}{\mathcal{N}}\left[(r / b)^3-3(a / b)(r / b)^2 R_1((b-r) / a)\right. \\
& +6(a / b)^2(r / b) R_2((b-r) / a)-6(a / b)^3 \\
& \left.\times R_3((b-r) / a)+6(a / b)^3 R_3(b / a)\right] \tag{12}
\end{align*}
for $r \leq b$ and
\begin{align*}
F_{mg}(r)= & 1-\frac{1}{\mathcal{N}}\left[3(a / b)(r / b)^2 R_1((r-b) / a)\right. \\
& \left.+6(a / b)^2(r / b) R_2((r-b) / a)\right] \tag{13}
\end{align*}
for $r>b$. The function is defined piecewise for \(r\leq b\) and \(r>b\), with normalization factor
\begin{equation}
\mathcal{N}=1+\left(\frac{a}{b}\right)^2\pi^2+6\left(\frac{a}{b}\right)^3R_3(b/a),
\end{equation}
where
\begin{equation}
R_k(x)=\sum_{n=1}^{\infty}(-1)^{n-1}\frac{e^{-nx}}{n^k}.
\end{equation}

In the above expressions, \(b\) is the half-charge radius and \(a=2.3/[4\ln(3)]\) characterizes the nuclear skin thickness. The parameter \(b\) is obtained from
\begin{equation}
b=\sqrt{\frac{5}{3}r_{\mathrm{rms}}^2-\frac{7}{3}a^2\pi^2},
\end{equation}
where the root-mean-square nuclear radius is estimated semiempirically as
\begin{equation}
r_{\mathrm{rms}}=\left(0.836A^{1/3}+0.57\right)\,\mathrm{fm},
\end{equation}
with \(A\) denoting the atomic mass number. 

\begin{table*}[t]
\caption{Individual contributions from various RCC terms and normalization factors of the RCCSDT method to the $A_{hf}$ constants (in MHz) of the considered states. $c.c.$ stands for complex conjugate terms and `Others' means the remaining RCCSDT terms that are not listed explicitly in the table.}
\begin{tabular}{lccccccccccc}
\hline \hline RCC terms & $3 S_{1 / 2}$ & $3 P_{1 / 2}$ & $3 P_{3 / 2}$ & $4 S_{1 / 2}$ & $3 D_{3 / 2}$ &  $3 D_{5 / 2}$ & $4 P_{1 / 2}$ & $4 P_{3 / 2}$ & $5 S_{1 / 2}$ &  $4 D_{3 / 2}$ & $4 D_{5 / 2}$ \\ 
\hline & & & & & & \\
$O$ & 623.97 & 63.43 & 12.60 & 150.56 & 0.59  & 0.25 & 20.98 & 4.17 & 58.16 & 0.25 & 0.11 \\
$\Bar{O}-O$ & 14.35 & 2.10 & 0.47 & 3.00 & $\sim 0$ & $\sim 0$ & 0.63 & 0.14 & 1.17 & $\sim 0$ & $\sim 0$ \\
$OT_1$ & 16.37 & 2.32 & 0.46 & 3.59 & 0 & 0 & 0.72 & 0.14 & 1.34 & 0 & 0 \\
$O S_{1 v}+$ c.c. & 105.16 & 10.18 & 2.01 & 18.21 & 0.02 & 0.01 & 2.71 & 0.54 & 6.01 & 0.01 & $\sim 0$ \\
$O S_{2 v}+$ c.c. & 128.81 & 17.10 & 3.16 & 29.86 & $-0.07$ & $-0.13$ & 5.12 & 0.97 & 10.99 & $-0.04$ & $-0.07$\\
$T_2^{\dagger} O T_{2}$ & 3.95 & 0.63 & 0.09 & 0.93 & $\sim 0$ & $\sim 0$ & 0.19 & 0.02 & 0.90 & $\sim 0$ & $\sim 0$\\
$S_{1 v}^{\dagger} O S_{1 v}$ & 4.43 & 0.43 & 0.07 & 0.55 & $\sim 0$ & $\sim 0$ & 0.09 & 0.02 & 0.16 & $\sim 0$ & $\sim 0$ \\
$S_{1 v}^{\dagger} O S_{2 v}+$ c.c. & 8.46 & 0.99 & 0.09 & 1.17 & $\sim 0$ & $\sim 0$  & 0.21 & 0.01 & 0.35 & $\sim 0$ & $\sim 0$ \\
$T_2^{\dagger} O S_{2v}$ & $-7.41$ & $-0.77$ & $-0.15$ & $-1.75$ & 0 & 0 & $-0.24$ & $-0.14$ & $-0.63$ & 0 & 0 \\
$S_{2 v}^{\hat{\dagger}} O S_{2 v}+$ c.c. & 11.29 & 0.93 & 0.22 & 2.70 & 0.01 & $\sim 0$ & 0.31  & 0.07 & 0.92 & $\sim 0$ & $\sim 0$\\
$S_{2v}^{\dagger}O S_{3 v}+$ c.c. & 1.49 & 0.14 & 0.03 &  0.36 & $\sim 0$  & $\sim 0$ & 0.05 & $\sim 0$ & 0.14 & $\sim 0$ & $\sim 0$ \\
$T_2^{\dagger}O S_{3 v}+$ c.c. & 6.87 & 0.50 & 0.20 & 1.64 & $\sim 0$ & $\sim 0$ &  $-0.07$ & $\sim 0$ & 0.54 & $\sim 0$ & $\sim 0$ \\
Others & $-15.59$ & $-1.47$ & $-0.08$ & $-3.56$ & $-0.02$ & $-0.02$ & 0.46 & 0.12 & $-1.61$ & $\sim 0$ & $-0.02$\\
Nm & $-3.60$ & $-0.15$  & $-0.03$ & $-0.53$ & $\sim 0$ & $\sim 0$ & $-0.04$ & $-0.01$ & $-0.18$ & $\sim 0$ & $\sim 0$\\
\hline
\end{tabular}
\label{terms_contri}
\end{table*}

\section{Results and discussion}

Before analyzing the hyperfine structure constants, it is essential to assess the accuracy of the atomic wave functions of Na obtained using the methods employed in their evaluation. To this end, Table~\ref{tab2} presents the IPs of the investigated $nS$, $nP$, and $nD$ states of Na calculated using the DHF, RMBPT(2), RCCSD, RCCSDTv, and RCCSDT methods. The estimated contributions from the Breit, VP, and SE corrections are also included, along with the experimental values from the National Institute of Standards and Technology (NIST) database \cite{kramida2018nist}. Comparison of the calculated IPs shows that electron-correlation effects exhibit a similar qualitative trend for all the states, differing primarily in their magnitudes. It is worth noting that the RMBPT(3) calculations employ the energies obtained at the RMBPT(2) level, whereas the RPA and other estimated corrections to the hyperfine structure constants are evaluated using DHF energies. As seen from Table~\ref{tab2}, the inclusion of electron-correlation effects progressively increases the IPs from the DHF to the RCCSDTv level. However, at the RCCSDT level, the additional treatment of higher-order core-correlation effects leads to a slight reduction in the IPs, resulting in closer agreement with the experimental values.

The Breit, VP, and SE corrections are relatively small and have only a marginal impact on the overall accuracy of the calculated IPs. The differences between the RCCSDTv and RCCSDT results indicate that higher-order core correlation effects are non-negligible, although their overall contributions remain modest. Our final recommended values are obtained by adding the Breit and QED corrections to the RCCSDT results. The quoted uncertainties include estimated contributions from the neglected high-lying virtual orbitals and from higher-order excitations beyond the RCCSDT approximation that could not be incorporated because of computational limitations. The excellent agreement between the calculated and experimental IPs, with discrepancies below 0.04\% for all the investigated states, demonstrates the high accuracy of the RCCSDT wave functions. This level of agreement provides strong confidence in the reliability of these wave functions for the subsequent evaluation of the hyperfine structure constants.

The calculated and available experimental values of $A_{hf}$ for the investigated states of $^{23}$Na are presented in Table~\ref{Ahf}. Similar to the IPs, the overall trend of the electron-correlation effects is consistent across all the states. However, unlike the IPs, the $A_{hf}$ constants exhibit a much stronger sensitivity to electron correlation. This behavior is expected because the magnetic dipole hyperfine interaction is governed by an operator that is highly localized in the nuclear region and therefore depends critically on the near-nuclear behavior of the electronic wave function. As expected, the DHF values decrease systematically from the $S$- to the $D$-states, reflecting the progressively smaller overlap of the valence-electron wave functions with the nucleus. Consequently, the strength of the hyperfine interaction decreases with increasing orbital angular momentum. Notably, the DHF value of $A_{hf}$ for the ground state is underestimated by nearly 30\% relative to the experimental value, underscoring the crucial role of electron correlation in accurately describing the hyperfine interaction. The RMBPT(2) and RMBPT(3) results are significantly larger than the DHF values and exhibit a monotonic increase, indicating that higher-order electron-correlation effects make substantial contributions to the $A_{hf}$ constants.

The RPA, which accounts for CP effects to all-orders, yields $A_{hf}$ values that lie between the RMBPT(2) and RMBPT(3) results. Since the RMBPT(2) method includes only the lowest-order CP contributions, this observation indicates that higher-order CP effects are relatively modest. In contrast, the BO and SR effects, which first appear at the RMBPT(3) method, play important roles in obtaining accurate $A_{hf}$ values. The RPA+BO results are consistently larger than the corresponding RMBPT(3) values. Because the third-order BO contributions included in the RPA+BO approach are also incorporated in the RMBPT(3) method, the smaller RMBPT(3) values suggest that the SR contributions partially cancel the BO contributions. This interpretation is further supported by the differences between the RPA+BO and RPA+BO+SR results. Finally, the differences between the RPA+BO+SR and RPA+BO+SR+Nm results are found to be negligible, indicating that the wave function normalization correction makes only a marginal contribution to the final $A_{hf}$ constants.

\begin{figure*}[t]
\centering
\includegraphics[width=16.0cm, height=14cm]{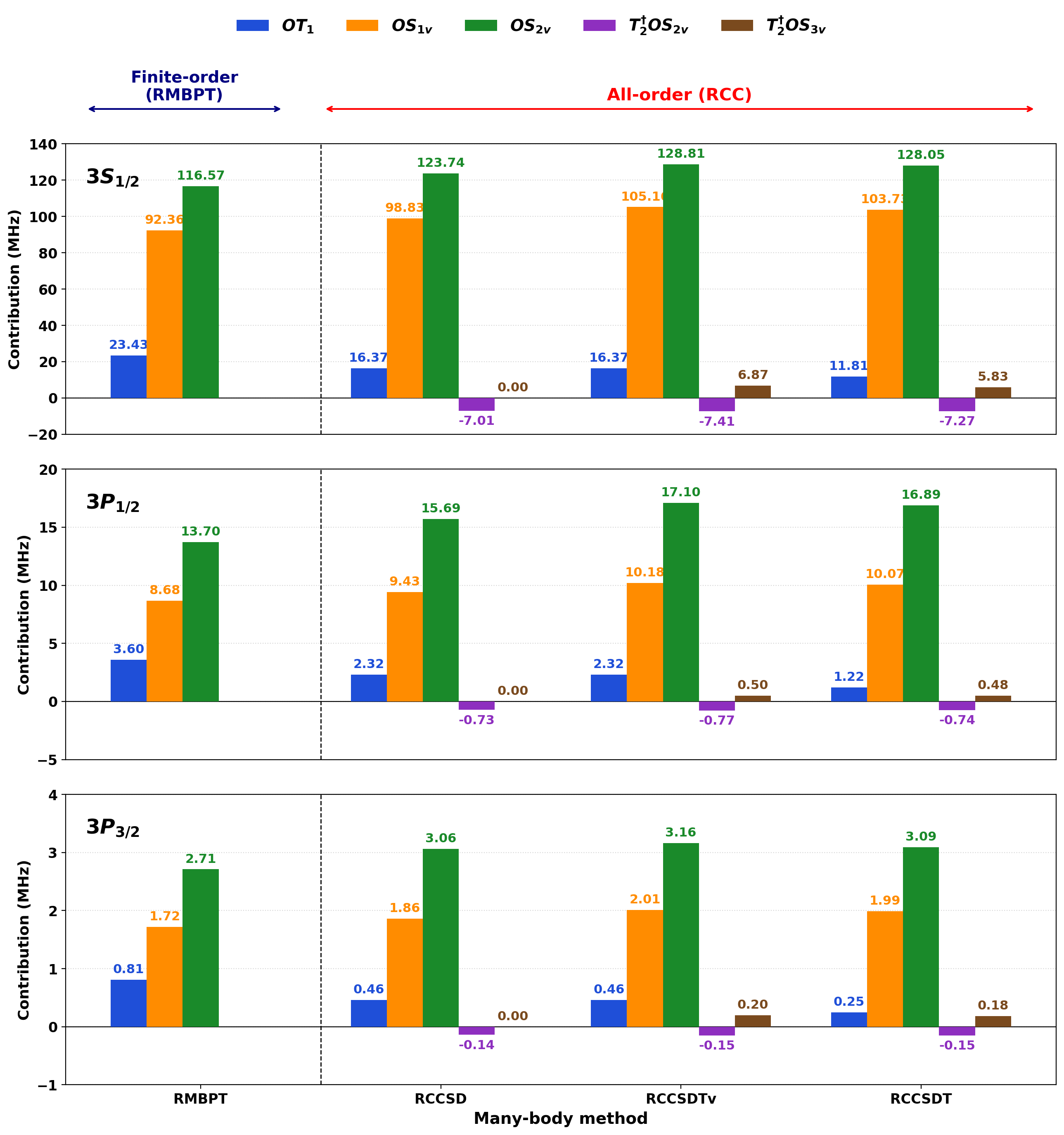}
\caption{Pictorial depiction of correlation contributions from the $OT_1+c.c.$, $OS_{1v}+c.c.$, $OS_{2v}+c.c.$, $T_2^{\dagger}OS_{2v}+c.c.$ and $T_2^{\dagger}OS_{3v}+c.c.$ terms to the $A_{hf}$ values of the $3S$, $3P_{1/2}$ and $3P_{3/2}$ states at the lowest-order (from the RMBPT method) and all-order methods with the RCCSD, RCCSDTv and RCCSDT approximations in $^{23}$Na (in MHz). It demonstrates influence of triples effects both directly and indirectly to the states.}
\label{evol_ahf}
\end{figure*}

The above discussion provides a qualitative understanding of the roles played by the different electron-correlation effects in determining the $A_{hf}$ constants of $^{23}$Na. The noticeable discrepancies between the RPA+BO+SR+Nm results and the experimental values indicate that higher-order PC and SR effects, together with their many-body couplings, are essential for achieving high-precision results. As discussed earlier, the RCC theory accounts for all these correlation effects to all-orders while also incorporating their mutual couplings in a unified framework. In the approximated RCC treatments, namely RCCSD, RCCSDTv, and RCCSDT, CP effects are included to all-orders, analogous to the RPA approach. In addition, these methods incorporate PC effects to all-orders and account for different levels of coupling among the CP, PC, and SR contributions. Among these approximations, the RCCSD values are consistently lower than the available experimental results, whereas the RCCSDTv values tend to overestimate them, demonstrating that valence triple excitations make significant contributions to the $A_{hf}$ constants in $^{23}$Na. A comparison of the RCCSDTv and RCCSDT results further reveals that the contributions from core triple excitations partially cancel those arising from the valence triples, bringing the RCCSDT values into much closer agreement with experiment. For the ground state, for example, the net contribution from the valence triple excitations exceeds 8 MHz, corresponding to more than 1\% of the DHF value. Similar trends are observed for the other low-lying $S$- and $P_{1/2}$ states. Since many high-precision applications require atomic-structure calculations with uncertainties below 1\%, often approaching 0.5\% or better, neglecting triple excitations would introduce systematic errors comparable to or even exceeding the desired accuracy. 

Figure~\ref{corr_trend} illustrates the systematic inclusion of different physical processes due to electron-correlation effects through lower- to higher-order many-body methods. It is evident from the figure that the $S$ and $P$ states exhibit positive correlation corrections, reaching the final results about $35\%-55\%$ of the DHF values. In contrast, inclusion of correlation effects in the $D$ states lower the values than the DHF results indicating negative correlation corrections; exceeding about 70\% in the $D_{5/2}$ states. This may be due to the fact that the $d$-orbitals of the DHF method have less overlap with the nuclear region, and large contributions from the $s$- and $p$-orbitals arise through the correlation effects. As a result, even moderate many-body corrections produce large relative changes, leading to the pronounced negative correlation corrections as seen in the above figure. The figure also demonstrates that the correlation corrections do not evolve monotonically with higher-level of approximation in the many-body method. For several states, the inclusion of RPA contributions leads to noticeable deviations from the RMBPT(3) results, while the subsequent incorporation of BO and SR effects partially compensates these changes, as already discussed before. Although the RCCSDT method yields a substantial improvement, noticeable deviations from the experimental values still remain. However, after incorporating the Breit, VP, SE, and BW corrections, the final recommended values are brought into excellent agreement with experiment. Among these contributions, the SE correction is particularly significant for the low-lying states. These observations demonstrate that, in addition to an accurate treatment of electron correlation, the inclusion of Breit, QED, and BW corrections is essential for achieving high-precision predictions of the $A_{hf}$ constants.

To gain a more quantitative understanding of the role of electron correlation in determining the hyperfine structure constants, Table~\ref{terms_contri} presents the individual contributions of the RCCSDT terms to the $A_{hf}$ constants. The contribution labeled $O$ corresponds to the DHF value for each state, while the remaining terms represent the electron-correlation corrections. Among these, the $OS_{1v}+c.c.$ and $OS_{2v}+c.c.$ terms (where $c.c.$ denotes the complex conjugate) provide the dominant contributions, as they account for the all-order PC and CP effects, respectively. The higher-order terms, such as $S^{\dagger}_{1v}OS_{2v}+c.c.$, $S_{2v}^{\dagger}OS_{2v}+c.c.$, and $T_2^{\dagger}OT_2$, are individually much smaller than the leading contributions; however, their cumulative effect is non-negligible. Contributions involving the valence triple-excitation operator, namely $S_{2v}^{\dagger}OS_{3v}+c.c.$ and $T_2^{\dagger}OS_{3v}+c.c.$, are found to be most significant for the low-lying $S$- and $P_{1/2}$ states, while remaining contributions are small. In the $D$ state, the DHF contributions are predominant followed by the contributions arising through the $OS_{2v}+c.c.$ terms. In fact, the DHF values are largely canceled by the correlation contributions arising from the $OS_{2v}+c.c.$ terms in the $D_{5/2}$ states, leading to small final values.

In Fig.~\ref{evol_ahf}, we illustrate the influence of triple excitations on the $A_{hf}$ constants explicitly. For the demonstration purpose, we only consider contributions to the ground state and the two low-lying excited states, $3P_{1/2}$ and $3P_{3/2}$, which possess different total angular momenta, only from a few important terms. Specifically, we analyze the contributions from the $OT_1+c.c.$ terms, which represent PC effects arising from core excitations; the $OS_{1v}+c.c.$ terms, which account for PC effects associated with valence excitations; the $OS_{2v}+c.c.$ terms, which describe CP effects; and the $T_2^{\dagger}OS_{3v}+c.c.$  terms, which provide the dominant explicit contributions from valence triple excitations in the property expression given by Eq.~(\ref{prop_exp}). To elucidate the evolution of these correlation effects with increasing many-body order, we also present the corresponding lowest-order contributions obtained within the RMBPT approach. Furthermore, the contributions from the $T_2^{\dagger}OS_{2v}+c.c.$ terms are included to facilitate a direct comparison with the $T_2^{\dagger}OS_{3v}+c.c.$ contributions, thereby highlighting the significance of the $S_{3v}$ amplitudes. It implies that although the explicit contributions associated with the valence triple-excitation terms are relatively small, their overall impact extends well beyond these individual matrix elements through their coupling with lower-order cluster amplitudes in the RCC equations.

\begin{table}[t]
\caption{Calculated $B_{hf}$ values (in MHz) of the states with angular momentum $J> 1/2 $ in $^{23}$Na using $Q=0.104(1)$ b \cite{stone2005table} at various levels of many-body approximations. The final values and uncertainties are listed as mentioned earlier and they compared with the previous calculation using non-relativistic RCCSD (CCSD) method and available experimental value.}
\resizebox{0.5\textwidth}{!}{\begin{tabular}{lcccccc}
\hline \hline \\
Method &  $3 P_{3 / 2}$ & $3 D_{3 / 2}$ &  $3 D_{5 / 2}$ & $4 P_{3 / 2}$ &  $4 D_{3 / 2}$ & $4 D_{5 / 2}$ \\
\hline  & & & & & & \\
DHF &  1.64 & 0.03 & 0.04 & 0.54 &  0.01 & 0.02\\
RMBPT(2) &  2.26 & 0.10 & 0.14 & 0.72 &  0.04 & 0.06\\
RMBPT(3) &  2.42 & 0.09 & 0.13 & 0.77 &  0.04 & 0.05\\
RPA &  2.56 & 0.14 & 0.19 & 0.81 &  0.06 & 0.08 \\
RPA+BO &  2.79 & 0.14 & 0.20 & 0.88 &  0.06 & 0.08 \\
RPA+BO+SR &  2.70  & 0.13 & 0.19 & 0.85 &  0.05 & 0.08\\
RPA+BO+SR+Nm & 2.70  & 0.13 & 0.19 & 0.85 &  0.05 & 0.08\\
RCCSD &  2.71 & 0.12 & 0.17 & 0.85 & 0.05 & 0.07\\
RCCSDTv  & 2.82 & 0.13 & 0.19 & 0.88 &  0.05 & 0.08\\
RCCSDT   & 2.78 & 0.13 &  0.19 & 0.87 &  0.05 & 0.08\\
\hline \\
$+$Breit & $-0.01$ & $\sim 0$ & $\sim 0$ & $\sim 0$ & $\sim 0$ & $\sim 0$\\
$+$VP  &   $\sim 0$ & $\sim 0$ & $\sim 0$ & $\sim 0$ & $\sim 0$ & $\sim 0$ \\
$+$SE  &  $\sim 0$ & $\sim 0$ & $\sim 0$ & $\sim 0$ & $\sim 0$ & $\sim 0$ \\
% $+$BW  & \\
\hline \\
Final &  2.77(5) & 0.13(1) & 0.19 (2) & 0.87(2) &  0.05(1) & 0.08(1) \\
CCSD \cite{salomonson1991coupled} &  2.68   & & & & &\\
MCDHF \cite{jonsson1996large}  & 2.72 & & & & &   \\
\hline
\hline
\end{tabular}}]
\label{Bhf}
\end{table}

The absence of contributions from the $T_2^{\dagger}OS_{3v}+c.c.$ terms in the RCCSD method of the aforementioned figure suggests that they do not contribute at this approximation, but they appear  in the RCCSDTv or RCCSDT methods through the corresponding property expression. It should be emphasized, however, that the influence of triple excitations is not restricted to the explicit terms containing the $T_3$ or $S_{3v}$ operators. The inclusion of triple excitations in the RCCSDTv and RCCSDT methods also modifies the single- and double-excitation amplitudes through the self-consistent solution of the RCC equations. Consequently, triple excitations contribute to the $A_{hf}$ constants both directly, through terms such as $T_2^{\dagger}OS_{3v}+c.c.$, and indirectly, through the renormalization of the $T_1$, $S_{1v}$ and $S_{2v}$ amplitudes that enter the leading RCC contributions. To distinguish these direct and indirect effects, we display the contributions from the $OT_1+c.c.$, $OS_{1v}+c.c.$, $OS_{2v}+c.c.$, $T_2^{\dagger}OS_{2v}+c.c.$ and $T_2OS_{3v}+c.c.$ terms obtained at the lowest-order RMBPT level and within the all-order RCCSD, RCCSDTv, and RCCSDT approximations in Fig.~\ref{evol_ahf}. It is evident that the $OS_{1v}+c.c.$  and $OS_{2v}+c.c.$ contributions undergo appreciable changes when going from the RCCSD to the RCCSDTv approximation. Since the RCCSDTv method includes only valence triple excitations, these changes arise entirely from the coupling of the $S_{1v}$ and $S_{2v}$ amplitudes with the valence triple-excitation amplitudes. Similar modifications are observed in the $OT_1+c.c.$ contribution when the RCCSDT results are compared with those from RCCSDTv. Although the $OT_1+c.c.$ terms do not explicitly contain triple-excitation operators, its magnitude changes significantly because the $T_1$ amplitudes are themselves modified through their coupling with the $T_3$ amplitudes in the RCCSDT equations (see Eq.~(\ref{rccsdt})). These results clearly demonstrate that the principal influence of triple excitations on the $A_{hf}$ constants arises indirectly through the self-consistent modifications of the single- and double-excitation amplitudes, rather than through the relatively small explicit contributions involving the $S_{3v}$ and $T_3$ operators. This finding underscores the importance of treating triple excitations self-consistently within the RCC framework instead of introducing them perturbatively. Such a treatment is essential for obtaining reliable sub-percent predictions of hyperfine structure constants in $^{23}$Na and, more generally, the same trend is also expected to follow in other alkali-metal atoms.

Table~\ref{Bhf} presents the calculated $B_{hf}$ constants arising from the E2 hyperfine interaction. Comparison of the results shows that electron-correlation effects are less pronounced for the $B_{hf}$ constants than for the corresponding $A_{hf}$ constants. Unlike the $A_{hf}$ values, the RMBPT(2) and RPA results indicate that the $B_{hf}$ constants are dominated by CP effects. Furthermore, the RMBPT(3) values are consistently smaller than the corresponding RPA results, exhibiting a trend opposite to that observed for the $A_{hf}$ constants. This behavior suggests that the BO contributions enhance the correlation effects only modestly, while the higher-order CP contributions remain dominant over the PC effects. This interpretation is further supported by the relatively small differences between the RPA and RPA+BO results. The differences between the RPA+BO and RPA+BO+SR values further indicate that the SR contributions partially cancel the combined CP and PC effects, similar to the behavior observed for the $A_{hf}$ constants. Finally, the negligible differences between the RPA+BO+SR and RPA+BO+SR+Nm results demonstrate that the wave function normalization correction makes only a marginal contribution to the $B_{hf}$ constants.

It is observed that electron-correlation effects in the $B_{hf}$ values are appreciable for the $3P_{3/2}$ state, whereas they remain relatively small for the other states. For the $3P_{3/2}$ state, the $B_{hf}$ values obtained within the RCCSDTv approximation are enhanced compared to the RCCSD results, indicating the significant role of valence triple excitations in shaping the corresponding wave functions. In contrast, the RCCSDT results are slightly lower than the RCCSDTv values, suggesting that contributions from core triple excitations partially cancel those arising from valence triples. Contributions from other corrections, such as Breit and QED effects, are found to be negligibly small. To the best of our knowledge, no experimental values are available for comparison with the present $B_{hf}$ results; however, theoretical data from CCSD and MCDHF methods exist for the $3P_{3/2}$ state of $^{23}$Na only \cite{salomonson1991coupled, jonsson1996large} and are included in Table \ref{Bhf}. A comparison shows that the present RCCSDT results are slightly larger than those obtained from previous calculations.

\section{Summary}

Ionization potentials and hyperfine structure constants of several states of sodium ($^{23}$Na), arising from magnetic dipole and electric quadrupole interactions, are investigated using a range of relativistic many-body methods. It is observed that electron-correlation effects lead to a systematic increase in the ionization potentials as one progresses from lower- to higher-order many-body approximations. In particular, triple excitations—especially those associated with valence correlations—play a significant role in achieving high accuracy. Contributions from Breit and QED effects are found to be negligible for these quantities. Nevertheless, the excellent agreement between the final theoretical results and available experimental data demonstrates the reliability of the atomic wave functions obtained using relativistic coupled-cluster methods.

Compared to the ionization potentials, electron-correlation effects are found to play a more pronounced role in the evaluation of the magnetic dipole hyperfine structure constants. Both core-polarization effects, treated within the random-phase approximation, and pair-correlation effects arising from Br\"uckner orbitals are found to be equally important for achieving accurate results. In contrast, structural-radiation contributions are relatively small and tend to contribute with an opposite sign compared to the dominant correlation effects. Within the relativistic coupled-cluster framework, the singles and doubles approximation yields lower values, while the inclusion of triple excitations systematically increases the results, bringing them closer to the available experimental data. In this context, higher-order relativistic corrections, including Breit and quantum electrodynamics effects, play a decisive role in achieving quantitative agreement with experiment. Contributions from triple excitations through core orbitals are also found to be significant, particularly for the ground state. The Bohr–Weisskopf effect is observed to be comparatively small for these properties.

In contrast to the magnetic dipole case, the accurate determination of electric quadrupole hyperfine structure constants is predominantly governed by an accurate treatment of electron correlation, especially core-polarization effects. Inclusion of pair-correlation effects further improves the precision of the results. While Breit and quantum electrodynamics contributions remain relatively small, correlation effects arising from triple excitations within the relativistic coupled-cluster framework are found to be essential and account for the differences between the present results and previously reported values.

\acknowledgments

B.K.S. is supported by the ANRF under grant no. CRG/2023/002558 and by the Department of Space, Government of India. All calculations reported in this work were performed on the ParamVikram-1000 HPC cluster at the Physical Research Laboratory, Ahmedabad, Gujarat, India.

\bibliography{references}
\bibliographystyle{IEEEtran}
\end{document}